\pdfoutput=1

\documentclass{iau}
\usepackage{graphicx}

\usepackage{amsmath}
\usepackage{amssymb}

\title[] 
{Waldmeier relations and the solar cycle dynamics by the mean-field dynamos}

\author{V.V. Pipin$^1$, D.D. Sokoloff$^2$ \and I.G. Usoskin$^3$}

\affiliation{$^1$Institute of Solar-Terrestrial Physics, Russian Academy of
Sciences, \\ Irkutsk, 664033, Russia,  email: {\tt pip@iszf.irk.ru} \\
$^2$Dept. of Physics, Moscow State University, \\ Moscow, 119991, Russia, email: {\tt sokoloff.dd@gmail.com} \\
$^3$Department of Physics, University of Oulu, \\
Oulu, 900014, Finland, email: {\tt ilya.usoskin@oulu.fi}}

\pubyear{2013}
\volume{294}  

\pagerange{--}

\jname{Solar and Astrophysical Dynamos and Magnetic
Activity}
\editors{A.G. Kosovichev, E.M. de Gouveia Dal Pino, \& Y.Yan, eds.}

\begin{document}

\maketitle
\begin{abstract}
The long-term variability of the sunspot cycle, as
recorded by the Wolf numbers, are imprinted in different kinds of
statistical relations which relate the cycle amplitudes,
duration and shapes.
This subject always gets a special attention
because  it is important for the solar activity forecast. We discuss  statistical
properties of the mean-field dynamo model with the fluctuating
$\alpha$-effect.
Also, we estimate dynamical properties of the model for
the long and short time-scale and compare it with the dynamics of the
sunspot numbers data sets.
\end{abstract}

\begin{figure}[ht]
\begin{centering}
\includegraphics[width=5.in]{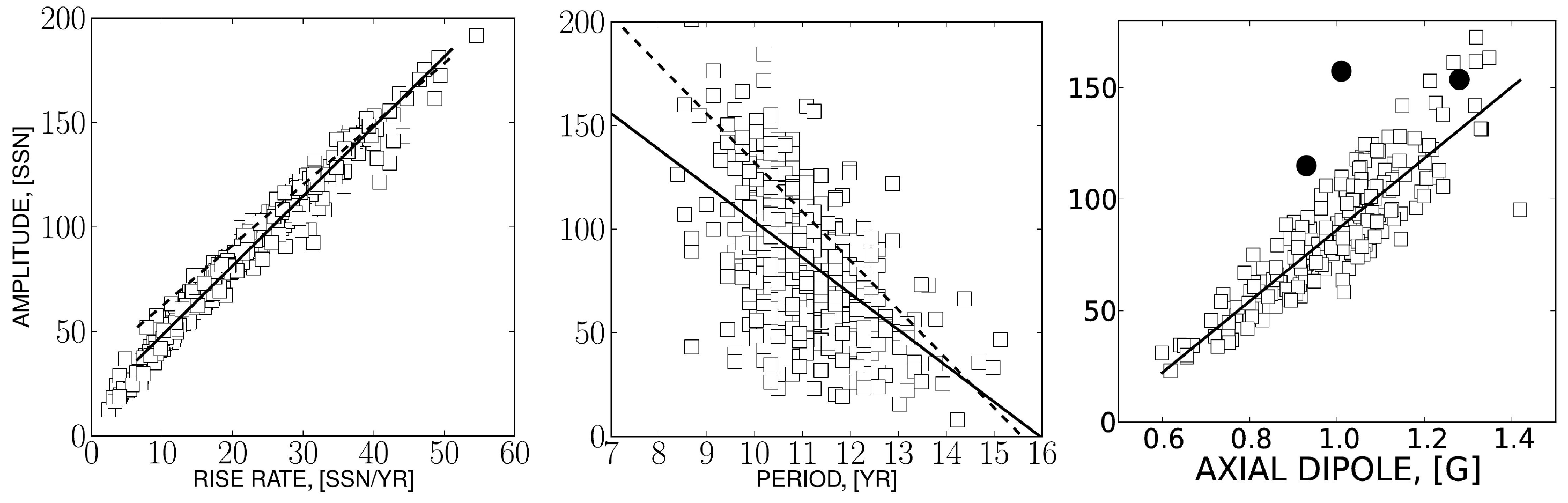}
\par\end{centering}
\caption{The Waldmeier relations for the model (the left and the
  middle panel); squares show data for individual cycles,
 while the solid line gives the correlation, the dashed
line shows these relations obtained from the actual sunspot
number(SSN) data. The circles  on the right panel 
  mark the results of the  Wilcox Solar Observatory (WSO) observations.
 } \label{GOR}
\end{figure}
\section{Waldmeier relations by the mean-field dynamos}

Recently,
Pipin \& Kosovichev (2011) showed that the Waldmeier relations can be
explained by the mean field dynamo model with a variable magnitude of
the $\alpha$-effect. The idea was elaborated further by Pipin \&
Sokoloff (2011) and Pipin et al. (2012) for dynamo models that take
into account the fluctuating $\alpha$-effect. Figure \ref{GOR} shows
some of the results obtained in those papers. Note, that the model reproduces the
Gnevyshev-Ohl rule as well (Pipin et al, 2012).
We suggest that  the obtained statistical relations
can be interpreted as an evidence for the solar cycle to be
a nonlinear self-excited oscillation that
tends to preserve the property of the attractor under random
perturbations.  The strength of the link between the
parameters of the subsequent cycles is controlled by the fluctuation amplitude
and by the perturbation's decrement. The latter strongly depends on the
nonlinear mechanisms involved in the dynamo.

\begin{figure}[ht]
\begin{center}
 \includegraphics[width=2.in]{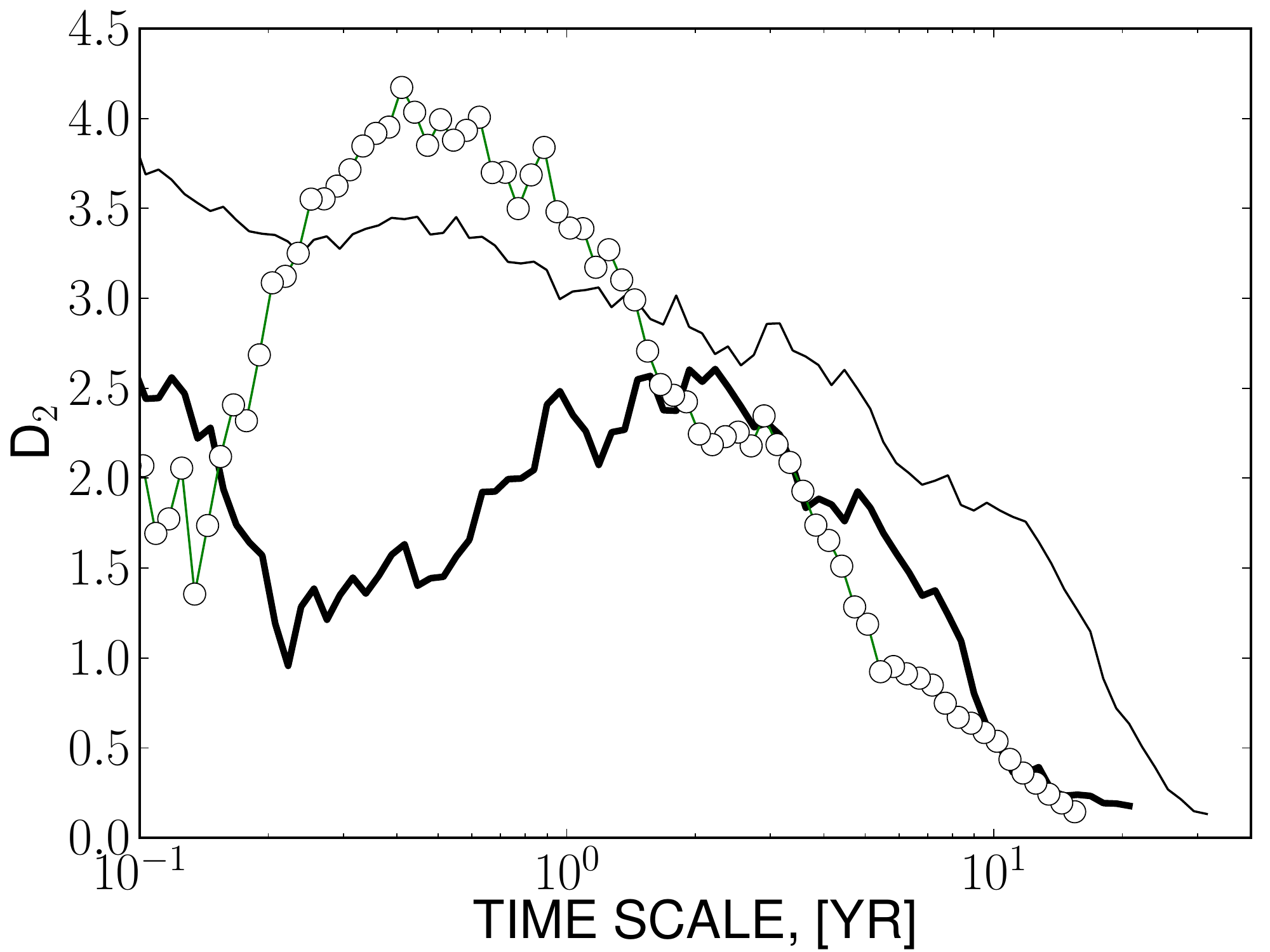}\includegraphics[width=2.in]{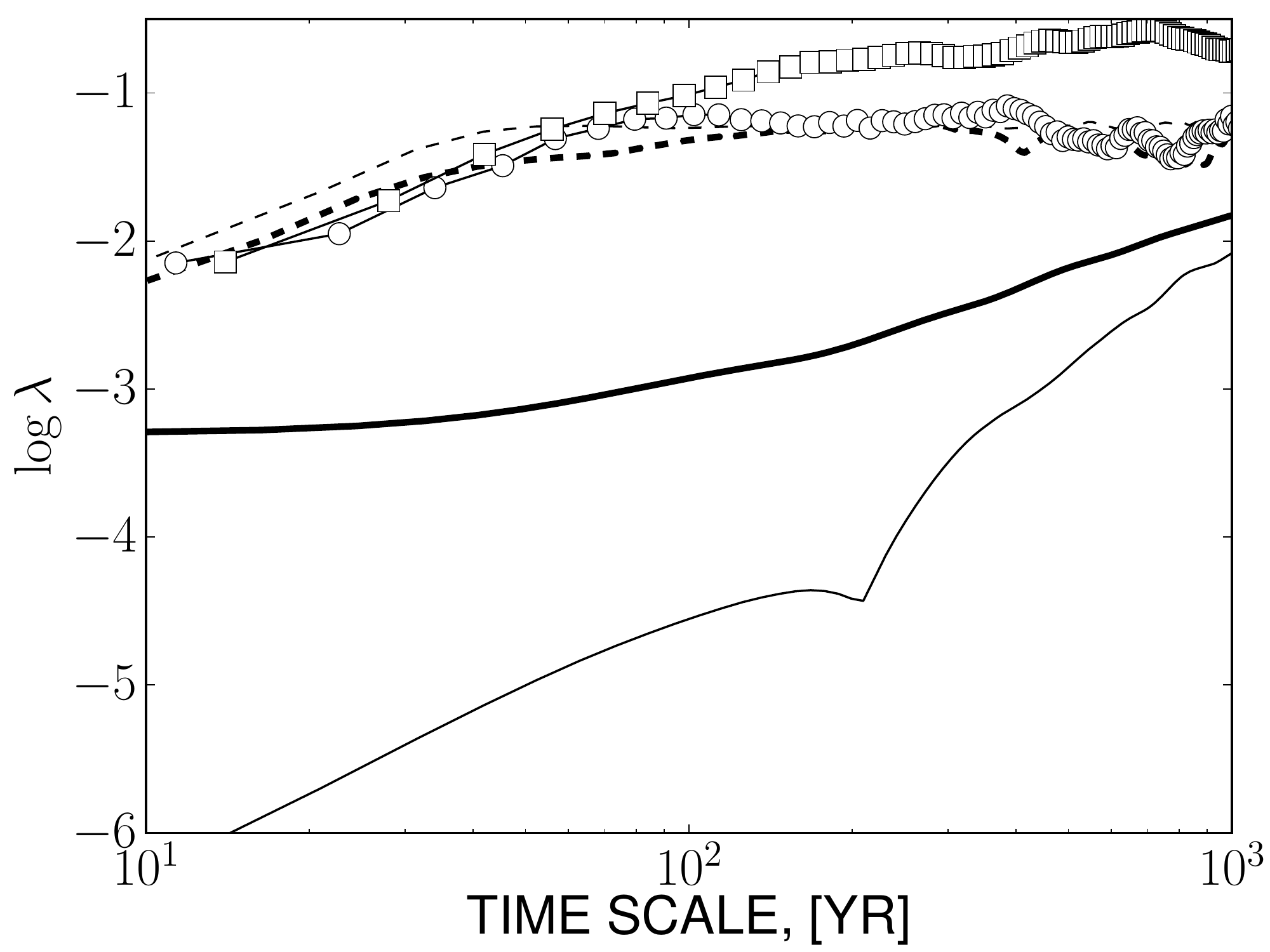}
 \caption{Left panel shows results for the correlation dimension, $D_2$,
   the solid thick line is for the 2D dynamo model, the thin line is
   for 1D dynamo model and the circles mark the results obtained from
   SSN data set. Right panel shows the same for the
   Lyapunov exponent spectrum. The squares mark the results for the 1D random walks
   model, the dashed curves show the results for the dynamo models
   data sets which were processed with  10 year running averaging.}
   \label{fig1}
\end{center}
\end{figure}

\section{The correlation dimension and Lyapunov exponents}
Bearing in mind the results discussed above, we tried to estimate
some of the nonlinear characteristics of the dynamo models, such as the
largest Lyapunov exponent, $\lambda$, and  the correlation dimension
$D_2$. The first one  shows the growth rate of deviations from a nominal cycle
with time (see, e.g., Kantz \& Schreiber 2004).
The correlation dimension $D_2$ quantifies the
topological complexity of the dynamical trajectories in the attractor.

Figure~2 (left) shows the correlation dimension $D_2$ for  the time series yielded by 
different types of nonlinear dynamo models (those include the
dynamical quenching due to magnetic helicity), including the 1D- and
2D models.  We observe that in all the data sets the value of
$D_2$ decreases to zero when the time scale approach the period of the
cycle.  The internal dynamics of 1D model (which anyway fails to reproduce the
Waldmeier relations) differs from the 2D case.
This is confirmed  by the results  for the  Lyapunov
exponent spectrum which
are shown at the right panel in Figure~2. The synthetic SSN data
(Solanki et al, 2004) shows
no exponential dynamics on the long-time scale. Dynamo models have it on
the short-time scale but lose on the long-time scale. This is
resulted from the different time resolutions of the data sets.
For comparison we put the spectrum for the simulated the 1D Brownian
motion process. This lead us to conclusion that on the long time scale the sunspot activity looks like
a random walk. The same behaviour is caught by the nonlinear
dynamo models with fluctuations.

{\bf Acknowledgments.}
VVP and DDS  thank for the support the RFBR grants
12-02-00170-a. VVP thanks  the support of the  Integration Project of SB RAS
N 34, and  support of the state contracts 02.740.11.0576, 16.518.11.7065 of the Ministry
 of Education and Science of Russian Federation.

\end{document}